\documentclass[letterpaper]{article}

\usepackage{natbib,alifeconf}  
\usepackage{algorithm}
\usepackage{algpseudocode}
\usepackage{amsmath}
\usepackage{xcolor}
\usepackage{hyperref}

\title{Spatial community structure impedes language amalgamation in a population-based iterated learning model}

\author{Georgia Sains, Conor Houghton$^*$ and Seth Bullock$^*$ \\
\mbox{}\\
Department of Computer Science, University of Bristol, Bristol, UK \\
$^*$equal contribution\\
\href{mailto:georgia.sains@bristol.ac.uk}{\texttt{georgia.sains@bristol.ac.uk}}} 

\begin{document}
\maketitle

\begin{abstract}
 The iterated learning model is an agent-based model of language evolution notable for demonstrating the emergence of compositional language. In its original form, it modelled language evolution along a single chain of teacher-pupil interactions; here we modify the model to allow more complex patterns of communication within a population and use the extended model to quantify the effect of within-community and between-community communication frequency on language development. We find that a small amount of between-community communication can lead to population-wide language convergence but that this global language amalgamation is more difficult to achieve when communities are spatially embedded. 
\end{abstract}

\section{Introduction}
Two of the defining features of human languages are, first, that they evolve and change through use and acquisition and, second, that they have a compositional syntactical structure that allows for the meaning of complex signals to be built out of the meanings of simple ones \citep{humanCommunication}. While other animals communicate, their signals are typically restricted to stereotypical cries that are inherited rather than learnt and few, perhaps even none, are able to manipulate or combine these signals as part of a compositional language \citep{animalsNoLanguage}. 

Iterated learning models (ILMs) are a family of computational models designed to simulate language change during acquisition in order to help demonstrate the conditions under which compositional languages come about \citep{basicILM}. In an ILM there is a language-using ``adult'' agent and a language-learning ``pupil'' agent. The adult teaches language to the pupil by exposing it to a finite series of utterances from which the pupil must infer the whole language. The pupil then becomes an adult and teaches a new pupil in the same fashion. A key feature of these models is their language learning bottleneck: since each adult only presents a finite subset of the language to their pupil, when the pupil in turn becomes an adult it will need to generate utterances that it has not experienced before. The process of generalization inherent to the model is thought to be critical to the behaviour of the ILM \citep{basicILM}.

In the original ILM, a single adult teaches a single pupil. While some variants of the ILM explore population settings for iterated language learning \citep{brace2015achieving,brace2016Understadning}, they are limited in that they treat adults and pupils as a single well-mixed community with no internal structure. In this paper, we propose an alternative population-based ILM where agents are nodes on a graph and communication takes place along the graph's edges. This allows exploration of the effect of population structure on language change. By placing agents within communities and varying the level of connectivity between them, we quantify how the nature and extent of community structure influences language emergence and evolution.

\section{The Iterated Learning Model (ILM)} 

In an ILM an agent is able to translate backwards and forwards between a set of meanings, $M$, and a set of signals, $S$. Each meaning represents some relationship between objects and ideas that are conveyed in language by a signal, while each signal can be understood to express one of the available meanings. For an agent, $i$, we will call their encoding map $e_i$:
\begin{equation}
    e_i:M\rightarrow S
\end{equation}
and the decoding map $d_i$:
\begin{equation}
    d_i:S\rightarrow M
\end{equation}
In the model, these two maps constitute an agent's language. Here, as in \citet{basicILM}, both the meanings and signals will be represented by 8-bit strings, which means both $M$ and $S$ are 256-element sets. In our model, $d_i$ is a neural network whereas $e_i$ is a lookup table which is generated through an \textsl{obverter} procedure defined below. 

Learning proceeds as follows. The teacher, $i$, picks a random subset of meanings and uses its lookup table encoder, $e_i$, to find the corresponding signals. This set of meaning-signal pairings is called the set of utterances and is communicated to the pupil. The pupil, $j$, will then use a number of randomised epochs of this set to train its own neural network decoder, $d_j$. Although the utterances only represent a subset of $M$; after training the pupil will obvert its current mapping to produce its own lookup table encoder $e_j$. Obversion will allow it to map every possible meaning to a preferred signal; though, of course, there is no guarantee that this map will map each meaning to a unique signal or that this map will be identical to the map belonging to its teacher. However, under the right conditions, the ILM results in the transmission of a language mapping that is indeed fully expressive (a unique signal is employed for each meaning) and stable (adult and pupil share the same mappings).

\begin{figure}[t]
\begin{center}
\includegraphics[width=\columnwidth]{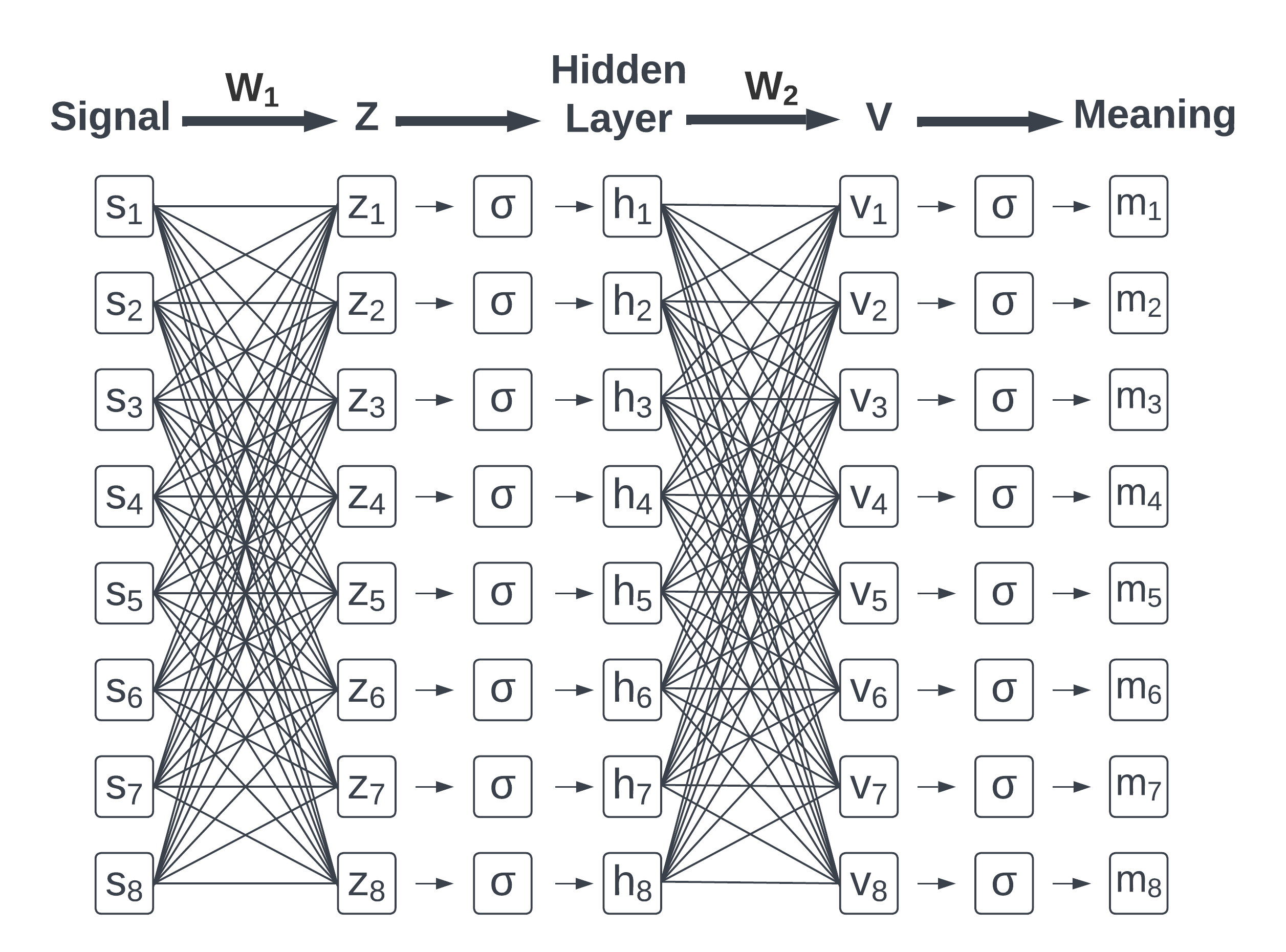}
\caption{The neural network decoder mapping from signal to meaning. The $Z$ and $V$ layers store the outputs of the network layers before passing them into the sigmoid function ($\sigma$). $W_1$ and $W_2$ are weight matrices, with $W_1$ comprised of independent weights mapping every character of the signal to every node of the $Z$ layer, and $W_2$ comprising independent weights mapping every node of the hidden layer to every node of the $V$ layer. }
\label{neuralNetwork}
\end{center}
\end{figure}

In our model, each $d_i$ is a fully-connected neural network (shown in Figure \ref{neuralNetwork}) with one 8-node hidden layer: 
\begin{align}
    \textbf{h}&=\sigma\cdot(W_1\textbf{s})\cr
    \textbf{m}&=\sigma\cdot(W_2\textbf{h})
\end{align}
 where signal $\textbf{s}$ is a binary 8-vector, the hidden layer $\textbf{h}$ is a real-valued 8-vector with components in $[0,1]$ and $\textbf{m}$ is a real-valued 8-vector, again with components in $[0,1]$, regarded as giving the probability that each meaning bit should be set to one. The nonlinearity $\sigma$ is a sigmoid, $\sigma (x) = 1/(1+e^{-x})$, and the dot in $\sigma\cdot(\textbf{x})$ indicates the non-linearity is applied component-wise. Thus the neural network associates a probability with the value of each bit in the output. The  meaning bit is taken to be a one if that probability is bigger than 0.5 and zero if it is less. The neural network uses stochastic gradient descent with no dropout and no bias terms. During learning, the agent's neural network is updated using backpropagation using the mean squared error (MSE) loss function with a learning rate of 0.1.
For an agent, $i$, the obverter procedure is used to create the lookup table for the map $e_i$ based on the mapping arising from $d_i$. While the agent principally uses $d_i$ to map from signals to meanings, it is a neural network which, in fact, produces a probability for each of the eight bits and can therefore be thought of as mapping from $S$ to a set of probabilities for each meaning:
\begin{eqnarray}
d_i^\prime&:&S\rightarrow(M\rightarrow[0,1])\cr
d_i^\prime&:&s\mapsto \{p(m;s)\mbox{ for }m\in M\}
\end{eqnarray}
where the probability of each meaning is calculated by multiplying the probability for each of its bits. This can be used to construct a full, $256\times 256$, table giving the probability for each mapping from any signal to any meaning. This table is called the \textsl{confidence table} and it must be calculated as part of the obversion procedure; a potentially computationally costly procedure if the number of bits were increased much beyond eight. To calculate $e_i$, each meaning is mapped to the signal with the highest probability in $i$'s confidence table:
\begin{equation}
    e_i(m)=\mbox{argmax}_s\,p(m;s)
\end{equation}
This is an example of a coordinated learning procedure to maximise communicative success \citep{obvert}. When trying to maximise the success of what signal to send, the obverter procedure considers what is the most likely signal to be understood by using the confidence table. Pseudocode for this procedure can be seen in Algorithm \ref{alg:obv}.

\begin{algorithm}[ht]
    \caption{Obverter Procedure Pseudocode}\label{alg:obv}
\begin{algorithmic}
\State $\mathrm{confidence} \gets 2^8\times 2^8$ array of 1's
\State mapping $\gets$ empty
\For{meaning $m \in$ space of possible 8-bit meanings, $M$}
    \For{signal $s \in$ space of possible 8-bit signals, $S$}
        \State $m' \gets \mathrm{NeuralNetwork.forward}(s)$
        \For{index $i$ in length($m$)}    
            \If{$m[i] = 0$}
                \State $\mathrm{confidence}[m][s] =$ \newline
                       \hspace*{15ex}$\mathrm{confidence}[m][s] \times (1-m'[i])$
            \ElsIf{$m[i] = 1$}
                \State $\mathrm{confidence}[m][s] = $\newline
                       \hspace*{15ex}$\mathrm{confidence}[m][s] \times (m'[i])$
            \EndIf
        \EndFor      
    \EndFor
    \State append $\langle m \mapsto \max(\mathrm{confidence}[m])\rangle$ to mapping
\EndFor
\end{algorithmic}
\end{algorithm}

In \citet{basicILM}, the languages that evolve from the ILM are quantified using measures of \emph{expressiveness} and \emph{stability}. Expressiveness is the proportion of the meaning space that is expressible from the adult's language. For a specific decoder map, $d$,
\begin{equation}
    \epsilon=\frac{|\{m:\exists{s}\mbox{ s.t. }m=d(s)\}|}{|M|}
\end{equation}
where $|S|$ is the number of elements in a set $S$.  Stability compares how similar two agent languages are in order to test how stable the language is after training. If $e_i$ and $e_j$ are the encoder mappings for the two agents, the stability is
\begin{equation}
    \sigma=\frac{|\{m:e_i(m)=e_j(m)\}|}{|M|}
\end{equation}
These metrics were used to compare the effect of different language learning bottleneck sizes. When the bottleneck is very small (20 utterances per generation) language does not stabilise between agents. When it is large (2000 utterances per generation), a stable and expressive language coalesces very quickly but has little compositionality. When the bottleneck is not too small but not too large (50 utterances per generation), a stable and expressive language develops that exhibits strong compositionality. This is because the bottleneck is large enough for a set of utterances to pass through it from which the relatively compact structure of a compositional language can be inferred, but is not so large that the less compact structure of a non-compositional language can be learned.

Compositionality, while not measured in \citet{basicILM}, is an important metric when considering ILMs as language evolution models. We use the definition for compositionality described in \citet{brace2015achieving}, where $C$ is equal to the average of the correlations between each bit of a language's meanings and the most correlated bit of the language's signals. We take all pairings of 8-bit meanings to 8-bit signals generated through an agent's obverter procedure and calculate the $8\times8=64$ correlations, $C_{xy}$, between the set of bits at location $x$ in the set of meanings and the set of bits at location $y$ within the associated set of signals to which these meanings map, for all values of $x$ and $y$. We then identify, for each value of $x$, the largest correlation coefficient, 
\begin{equation}
C^*_x = \max_{y}C_{xy}
\end{equation}
The average of these eight values gives an overall value for compositionality of the language
\begin{equation}
    C=\frac{1}{8}\sum_{x}C^*_x
\end{equation}

\begin{figure}[ht]
\begin{center}
\includegraphics[width=0.8\columnwidth]{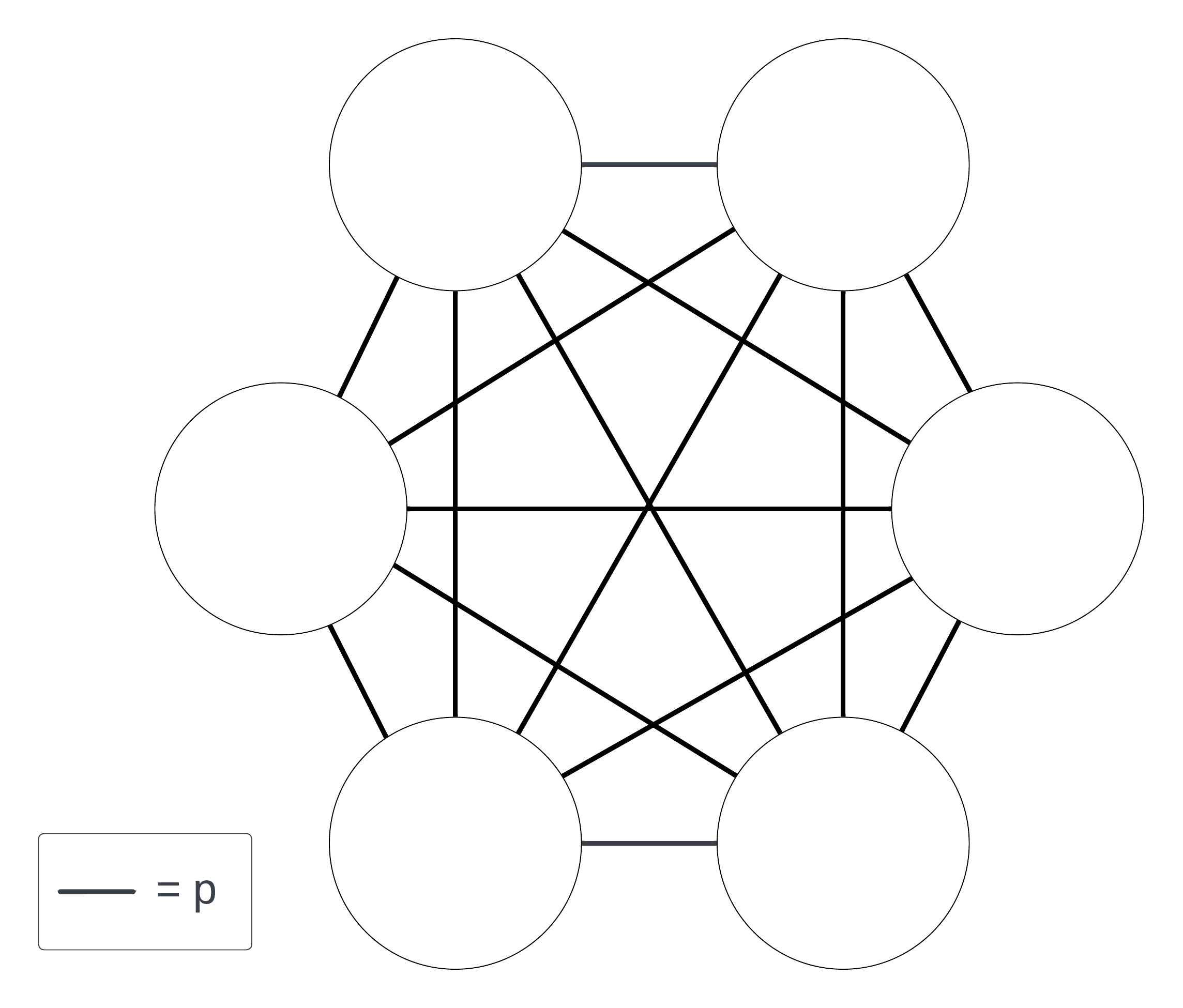}
\includegraphics[width=0.8\columnwidth]{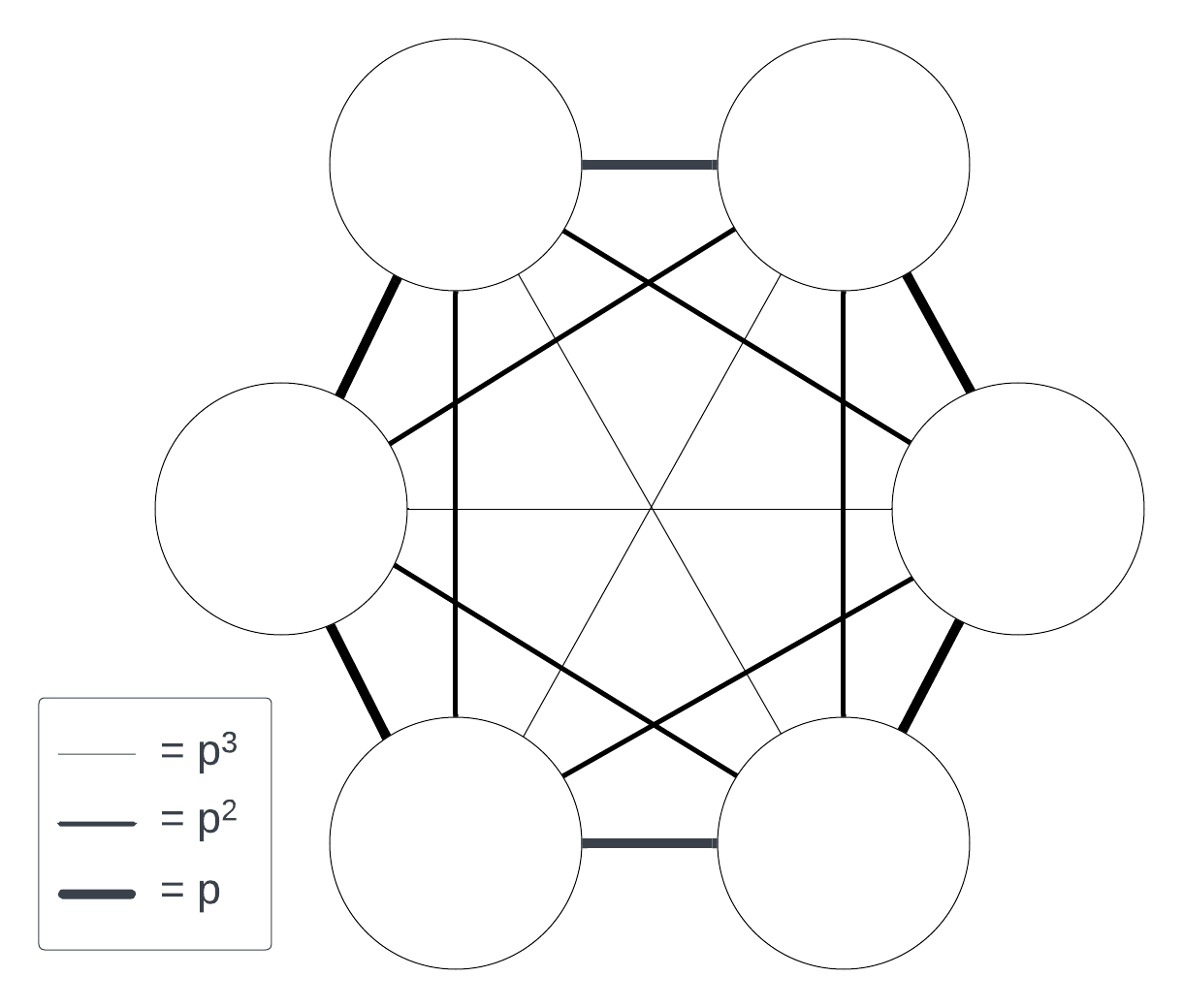}
\caption{Population structure for the unstructured community-based ILM (top) and the spatial community-based ILM (bottom). Each circle represents a community of $N$ agents, with the connection strength between agents from different communities being equal to $p$ in the unstructured case, and being equal to $p^{D}$ in the spatial case, where $D$ is the distance between any two communities.}
\label{fig:structure}
\end{center}
\end{figure}

\begin{figure*}[ht]
\begin{center}
\includegraphics[width=\textwidth]{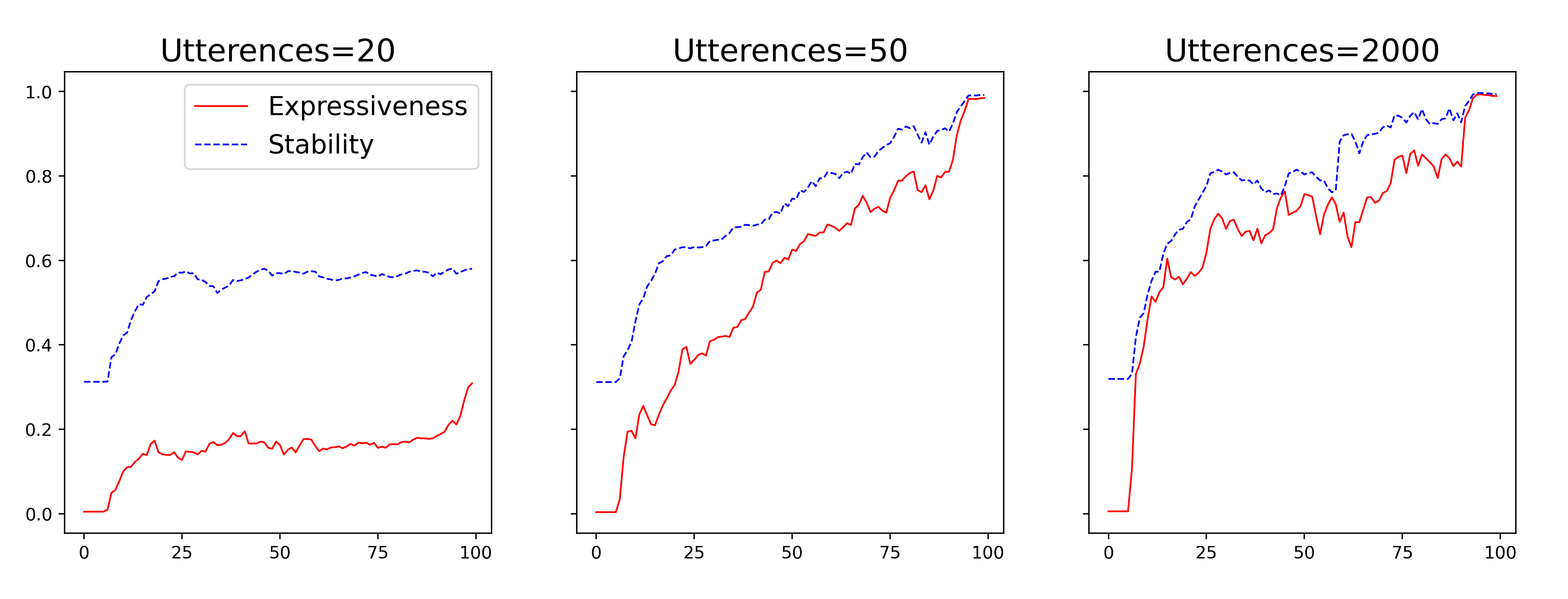}
\caption{The effect of different bottleneck sizes on the change in mean expressiveness and mean stability of language use during one 100-generation run of the population-based ILM for a population of 10 agents on a complete graph.}
\label{fig:utterances}
\end{center}
\end{figure*}

\begin{figure*}[ht]
\begin{center}
\includegraphics[width=\textwidth]{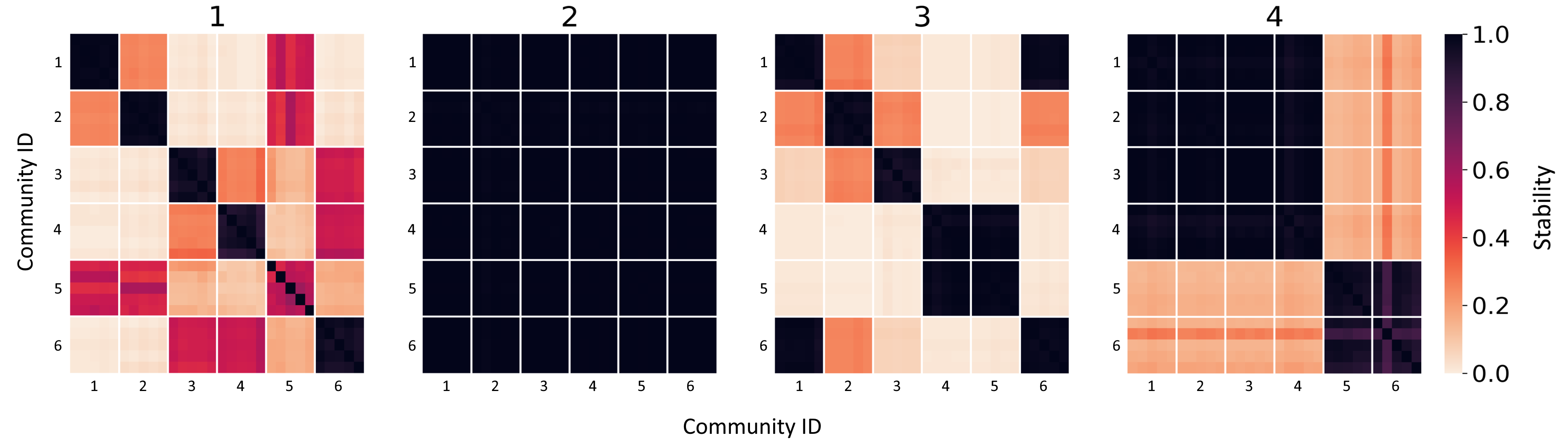}
\caption{Heatmaps showing four different example outcomes for the community-based ILM. All four simulations were run for 1000 generations with $G=6$ communities of $N=5$ agents, and agent replacement was disabled for the last 40 generations to allow all agents to stabilise. Each of the $30 \times 30$ cells in each heatmap shows the language stability between a pair of agents, with white lines showing the separation between communities. Heatmaps \textbf{1} and \textbf{2} show results for between-community communication rates of 15\% and 20\% in the unstructured community-based ILM, respectively. Heatmaps \textbf{3} and \textbf{4} show results for between-community communication rates of 15\% and 20\% in the spatial community-based ILM, respectively.   Within-community language stability corresponds to the leading diagonal blocks, whereas between-community stability corresponds to the off-diagonal blocks. For the spatial case, the distance of a block from the leading diagonal or from the top-right or bottom-left corner (whichever is closer) indicates the degree of spatial separation between the two communicating communities.} 
\label{fig:heatmap}
\end{center}
\end{figure*}

\section{Population-Based ILM}

In this paper, we explore two variants of the ILM. The first of which is a population-based ILM which introduces multiple concurrently existing agents on a fully-connected graph. The second, defined in the next section, is a community-based ILM which takes the population model and imposes some community structure on the graph.

The population-based ILM contains $N$ agents that are located on nodes within an interaction graph describing which agents learn language from which other agents. The network fully connects all pairs of nodes as a complete graph (excluding self-connections). Each agent is initially naive, with its language initialised by assigning random weights to every connection in its neural network. Each naive agent is also assigned a lookup table mapping meanings to signals which is initially empty.

The model proceeds by simulating a series of generations during which each agent gets older, and one agent teaches its language to all of the others. An ILM requires that pupils eventually replace the teachers that teach them and are themselves replaced by new naive agents that enter the population. In our population model agents are replaced in a staggered manner, with one adult agent being removed after each generation and replaced by a new naive agent with the same population network connections. Each incoming naive agent will become an adult after five generations. After these five generations, the naive agent has their empty lookup table populated with a mapping from every possible meaning to their preferred signal as generated by applying the obverter procedure to their neural network.

At the start of each generation, one adult, $T$, is selected uniformly at random to be a teacher. In this way each naive agent will learn from, on average, five different adults before it becomes an adult itself. In the first four generations there are no adults and as such no communication. After this however, the teacher communicates to all connected agents at a time ensuring each language-receiving agent will receive one communication per generation; the number of utterances the teacher passes to the adults who audit them is chosen so that on average, after five generations, a naive agent receives the target amount of utterances. This target amount of utterances is the variable that will be compared and will use the same values as \citet{basicILM} of 20, 50 and 2000. Since both naive agents and adults can receive language, adults also perform minor language acquisition when receiving language where they perform a greedy partial version of the ILMs language learning procedure. They still perform the backpropagation step, but the obverter procedure will only iterate over the pairs of meanings and signals witnessed during the current generation as opposed to the whole meaning space and signal space.

\section{Community-Based ILM}
Using this population-based ILM, we can now consider how agents learn in communities. The goal is to understand what level of communication outside of one's community is required for a population-wide shared language to arise. Additionally, we will explore the effect that spatial organisation of communities has on language amalgamation. Two models have been developed as variants of the Population-Based ILM, an \emph{unstructured} community-based ILM and a \emph{spatial} community-based ILM. 

For the unstructured community-based ILM, we can define a population as $G$ communities each comprising a complete graph of size $N$ and each connected to the other communities with some degree of connectivity, $p$. When an adult agent, $i$, communicates, for each agent $j$ outside their community there is a probability $p$ that $j$ will receive the communication from $i$, whereas each agent $k$ within the same community as $i$ will receive their communication with probability one. This structure can be seen in Figure~\ref{fig:structure}. The quantity of epochs of utterances sent with each communication step is reduced based on the number of pupil agents to maintain total language output per generation. This is because the number of pupil agents is variable in this model. We can calculate the expected total proportion of a teacher's communication that is between community ($BC$), as opposed to within their own community, as:
\begin{equation}
    BC = \frac{(G-1)Np}{(N-1)+(G-1)Np}
    \label{eq:p}
\end{equation}
We set the number of adults selected to be teachers each generation, $\mathcal{T}$, to be equal to the number of communities, $G$ so that on average each generation there is one teacher for every community in the model's population. The model uses a bottleneck of 50 utterances received for each agent per generation as this was shown by \citet{basicILM} to result in the development of compositional language. Additionally, the staggered agent replacement replaces agents from one community at a time. This prevents all agents from being replaced in a single community sequentially.


For the spatial community-based ILM, we arrange a population's communities on a ring similar to that of a connected-caveman graph \citep{connectedCaveman}. Here, rather than between-community communication between any pair of communities being equally likely, the probability of between-community communication is dependent on the ring distance between the communities as shown in Figure~\ref{fig:structure}. The probability of communication between agents of different communities with distance $D$ between communities is equal to $p^D$, with $p$ being the probability of communication of neighbouring communities. This means that communication occurs more frequently between communities that are spatially close to one another. The average rate of between-community communication, $BC$, can also be calculated for this spatial model:
\begin{equation}
    BC = \frac{\sum_{x=2}^{G}Np^{\lfloor x/2 \rfloor}}{(N-1)+\sum_{x=2}^{G}Np^{\lfloor x/2 \rfloor}}
    \label{eq:Sp}
\end{equation}
For both of these two models, language stability was measured for every pair of agents in the final population allowing us to assess the pattern of language agreement across the whole population and see which communities agree on a shared language. Additionally, we calculated two additional values, \emph{in-stability} and \emph{out-stability}. Out-stability is the average stability between all pairs of agents that are not members of the same community, while in-stability is the average stability between all pairs of agents that belong to the same community. The latter indicates the extent to which language evolution has been successful within each of a population's communities while the former indicates the extent to which the whole population shares a single language.

\section{Results}
First, the behaviour of the population-based ILM was explored for a population of size 10 on a complete graph (i.e., there was no community structure within the population). The impact of varying bottleneck size was assessed using the values from \citet{basicILM} of 20, 50 and 2000. The results can be seen in Figure~\ref{fig:utterances} which shows that the population-based ILM behaves similarly to the original ILM with one difference being that language stability and expressiveness change more gradually over generational time. This results from the gradual turnover of agents within the population. Whereas a teacher is replaced by their pupil in each generation of the original ILM, the population-based ILM requires that multiple generations must pass before all teachers are replaced by their pupils.

\begin{figure}[ht]
\begin{center}
\includegraphics[width=\columnwidth]{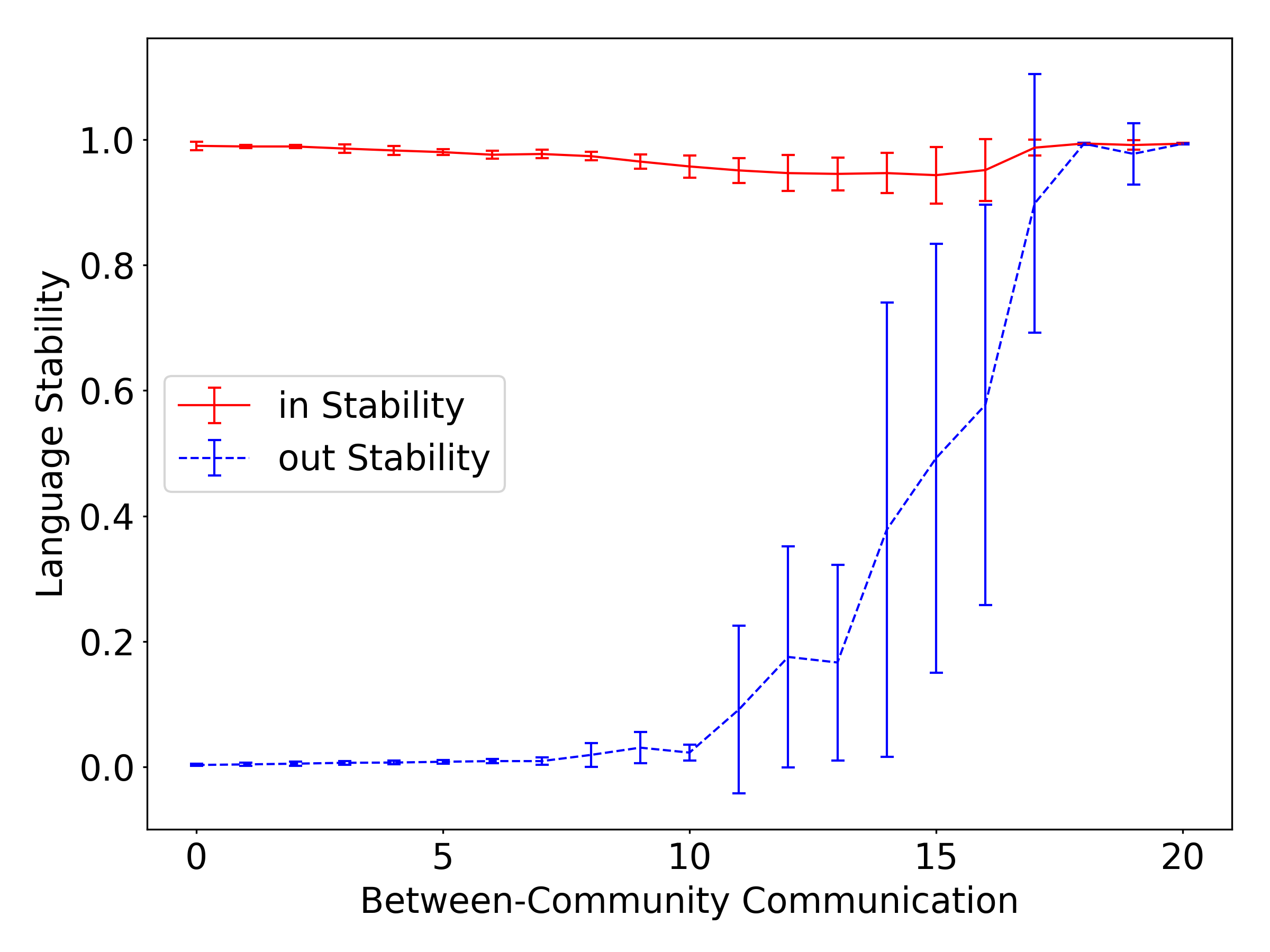}
\caption{Variation in the average final in-stability and out-stability of language use with the extent of between-community communication for the unstructured community-based ILM. Plots show the average values for ten simulations with error bars showing standard deviations. Simulations were run for 1000 generations with $G=6$ communities of $N=5$ agents. Agent replacement was disabled for the last 40 generations to allow all agents to stabilise.}
\label{fig:unstructured}
\end{center}
\end{figure}

Additionally, for a bottleneck size of 50, the language developed was completely compositional, with every digit of the meaning mapping to a unique digit of the signal. With a bottleneck of size 2000, the language developed was not compositional with $C=0.83$, which is low given the minimum possible value for $C$ is 0.5. This lack of compositionality is also indicated by the erratic rate at which the language changes. Large changes can happen quickly in this scenario as there is no structure to the language being developed. 

Example outcomes of language evolution for between-community communication rates of 15\% and 20\% can be seen in Figures~\ref{fig:heatmap}.\textbf{1} and \ref{fig:heatmap}.\textbf{2}, respectively. The effect of introducing communities to this model can be seen in Figure~\ref{fig:unstructured}. Communities stabilise quickly when the rate of between-community communication reaches 10\%, with consistent total stability achieved for $BC>18$\%. 

\begin{figure}[ht]
\begin{center}
\includegraphics[width=\columnwidth]{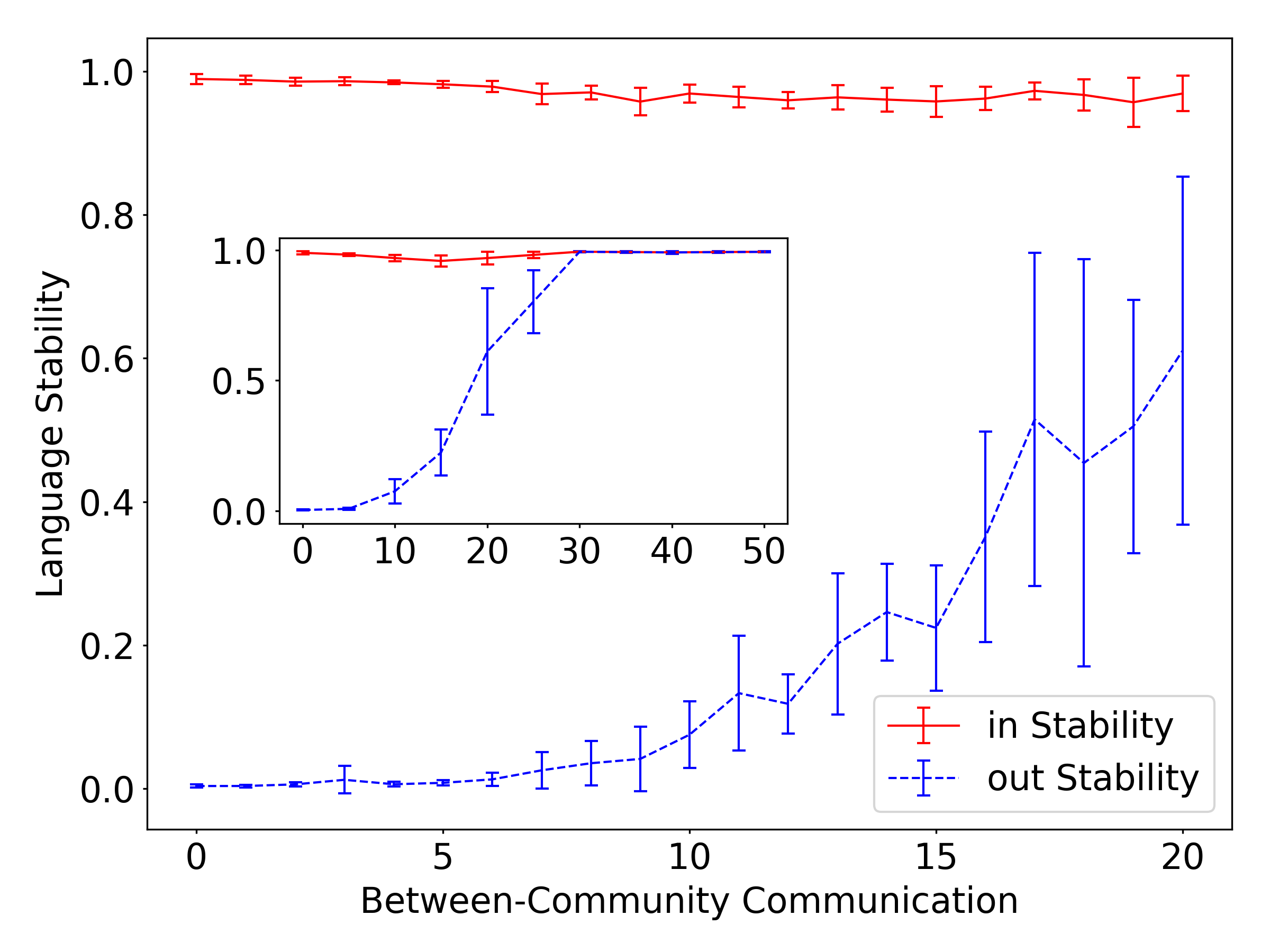}
\caption{The effect on in-stability and out-stability of varying the rate of between-community communication in the spatial community-based ILM. Plots show the average values for 10 independent simulations with error bars showing standard deviations. Simulations were run for 1000 generations with $G=6$ communities of $N=5$ agents. Agent replacement was disabled for the last 40 generations to allow all agents to stabilise. The additional inset graph shows the behaviour of this model up to 50\% external communication.}
\label{fig:spatial}
\end{center}
\end{figure}

\begin{figure*}[ht]
\begin{center}
\includegraphics[width=\textwidth]{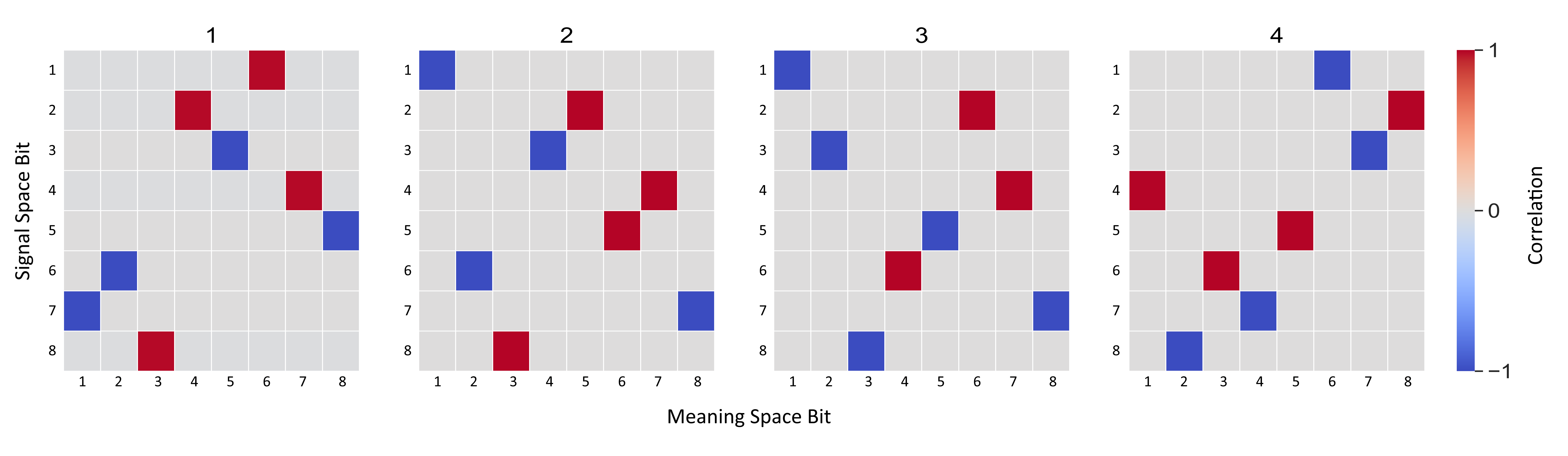}
\caption{The visualised language output of all languages from Figure \ref{fig:heatmap}.\textbf{3}. Each individual heatmap shows the mapping of meaning bits to signal bits for a language. Heatmap \textbf{1} shows the language shared by communities \textbf{1} and \textbf{6}. Heatmap \textbf{2} shows the language developed by community \textbf{2}. Heatmap \textbf{3} shows the language developed by community \textbf{3}. Heatmap \textbf{4} shows the language shared by communities \textbf{4} and \textbf{5}. A value close to zero means that there is little or no correlation for that bit between meaning and signal, a value of one means there is a perfect positive correlation (\textbf{1} maps to \textbf{1} and \textbf{0} maps to \textbf{0} for all 256 meanings) and a -1 means there is a negative correlation (\textbf{1} maps to \textbf{0} and \textbf{0} maps to \textbf{1} for all 256 meanings). In each language here all the correlations have values equal to, or nearly equal to one, zero or -1; this is because these languages are all compositional; it is not a inevitable property of languages. The languages for communities \textbf{1} and \textbf{2} have three bits in common, \textbf{2} and \textbf{3} have three bits in common, whereas \textbf{1} and \textbf{3} have one.} 
\label{fig:languageHeatmap}
\end{center}
\end{figure*}

When communities are structured spatially (Figure~\ref{fig:spatial}), population-wide language stability can be seen for rates of between-community communication as low as 6\%. However, stability increases more slowly with increasing between-community communication rates, only reaching fully stable populations at rates of 30\% or higher. Outcomes of language evolution for between-community communication rates of 15\% and 20\%  are shown in Figures~\ref{fig:heatmap}.\textbf{3} and \ref{fig:heatmap}.\textbf{4}, respectively. Our interest here is in stability and its interaction with community structure. We did measure expressiveness and compositionality but neither showed any marked difference from the non-spatial model.

An example of languages developed from the spatial model can be seen in Figure~\ref{fig:languageHeatmap}. This shows all four languages that were developed in the simulation from Figure~\ref{fig:heatmap}.\textbf{3}. All four languages are entirely compositional, as seen by the direct mappings of meaning bits to signal bits. Language \textbf{4} is entirely unique, while different numbers of overlapping bits are shared across languages \textbf{1}, \textbf{2} and \textbf{3}.

\section{Discussion}
The behaviour of the population-based ILM is consistent with that of the original ILM by \citet{basicILM}. While language changes more smoothly, the key result that learning bottleneck size determines language compositionality remains. Against this background, we can assess the impact of introducing community structure within the agent population.

For populations of agents divided into communities where interactions between any pair of unique communities are equally likely, language fully amalgamates across the whole population for relatively small rates of between-community communication ($\approx18$\%). For the population size that we consider here, population-wide language arises for rates of between-community communication of less than one ($\approx 0.88$) communication per agent per generation  This shows that the amount of interaction required between communities in order for languages to coalesce can be relatively small. Additionally, a population's in-stability is reduced for scenarios where the rate of between-community communication is high enough to cause the population to partially converge on a shared language. Small amounts of external communication can negatively impact some communities, causing interference that hinders their ability to converge on a shared language. An example of this can be seen in Figure~\ref{fig:heatmap}.\textbf{1}, where the fifth community has a language that is not completely internally stable within the community.

Introducing spatial community structure might be expected to allow population-wide language to emerge more easily as communication between adjacent communities would be more intense allowing them to more readily agree on a shared language. The opposite is in fact the case, with spatial community structure requiring considerably higher rates of between-community communication ($\approx 30$\%) before shared population-level language arises. While this may seem counter-intuitive, Figure~\ref{fig:heatmap}.\textbf{4} shows why this is the case. More frequent communication between adjacent communities leads different adjacent pairs of communities to settle on different shared languages simultaneously. While this allows some communities to develop the same language, it also leads to higher-order competition between alternative languages each of which is spoken by more than one community. Without high rates of between-community communication, these competing multi-community languages persist and do not merge, leading to a final state in which two or three different languages are maintained in the population.

This behaviour can be further explained when we consider compositionality. The language learning bottleneck requires languages to develop compositionality \citep{basicILM,brighton2002compositional,smith2009iterated}. Since $8!$ different possible compositional languages are available in this model, the structure of two languages arising simultaneously within the model will almost certainly be distinct from one another. Therefore, when multiple different compositional languages develop in the same population it takes a significant amount of communication between the different language communities before the two grammars tend to amalgamate, explaining the persistence of multiple languages within the spatial community-based model.

Like our model, the population-based ILM introduced in \citet{brace2015achieving} incorporates learning from multiple agents. However, it differs from ours in that it uses a strict generation structure. They showed that language can develop in an ILM with multiple teachers and some horizontal language transmission. In a sense, our goal was to build on this by demonstrating an ILM with a still more flexible interaction structure. In \citet{steels1998spatially} there is a discussion of a ``Naming Game'', a type of language game where agents attempt to agree upon names for objects. This is expanded by \citet{dall2006agreement} who used the Naming Game in small-world networks to explore convergence towards a global agreement using a parameter for rate of rewiring in the small-world network. They found that as this parameter increased, the rate of convergence greatly increased. Our model uses a similar network, but with quite different learning dynamics. Of course, our choice of network is intended as an example, but is not based on principle; in fact, the true dynamics of language evolution will involve a complex relationship between network and language and would involve different types of interaction between the agents. Indeed, the language might not only to support communication but also to act to create the network by acting as a signifier for in-group and out-group identity \citep{thomas2018self,benitez2020four,cambier2022prosociality}.

While the current model is effective in exploring the emergence of a single shared language in a population of language users, real-world language acquisition from multiple sources often leads to bilingualism \citep{bilingual} which is a phenomenon not present in this model. \citet{multipleLanguages} explores this in an ILM environment using Bayesian agents developed by \citet{griffiths2007language}. The agents maintain hypothesis distributions over multiple languages allowing them to adopt various degrees of bilinguality. However, one simplification of this model, as noted by \citet{brace2015achieving}, is its assumption of vertical language transmission. Exploring multilingual agents in the model presented in this paper (perhaps using a more subtle representation of bilinguality) would allow exploration of bilingual language formation in communities that initially speak different languages.

\section{Conclusion}
This paper presents two extensions to the iterated learning model described by \citet{basicILM}. The novel population-based model allows the dynamics of population turnover and social interaction on a network to be incorporated into the language evolution model. Introducing community structure within the language population demonstrated the relatively small rate of between-community communication that is required for a population-wide language to emerge across unstructured communities, and also allowed the effect of spatial community structure on language evolution to be explored. When communities were organised in a spatial ring configuration the discovery of a population-wide language was more challenging. While adjacent communities would tend to develop a shared language, these multi-community languages would compete at the population level and resist being amalgamated. Overall, while language can propagate across a networked population to some extent, spatially correlated community structure can prevent language convergence at the population level unless between-community communication rates are significant.
\section{Code availability}
\href{https://github.com/pop-ILM/ALIFE2023}{\texttt{https://github.com/pop-ILM/ALIFE2023}}

\section{Acknowledgements}
CH is supported by the Leverhulme Trust, grant number RF-2021-533.

\section{Appendix}
The parameters for both models are given here along with some explanation of why they were chosen.
\begin{itemize}
    \item \textbf{Signal Size \& Meaning Size:} A signal size and meaning size of eight bits was chosen as this matches the original ILM described by \citet{basicILM}. The structure of the neural network (with one hidden layer of eight nodes and a learning rate of 0.1) as well as the number of epochs (100) are taken from this as well.
    \item \textbf{Population size:} Ten agents were used in a complete-graph for the population-based ILM as we wanted a large number of agents to demonstrate the novel population interaction. However, we could not make the number too large because of the polynomial increase in number of edges with nodes in a complete graph and the computational cost of the model.
    \item \textbf{Community Structure:} Six communities of five agents were used in this paper. Six communities were chosen to further distinguish between the unstructured and spatial community-based ILM, as six communities means each community has an opposite community with a distance of three. Five agents were chosen as the community size as this kept the number of edges in the whole graph low enough to be run in reasonable time.
    \item \textbf{Number of generations:} 100 generations were used for the population-based ILM as this was shown in \citet{basicILM} to be enough generations to consolidate language. 1000 generations were used for the community-based ILM to give the population a large amount of excess time to stabilise the final languages even with the added influence from other communities.
\end{itemize}

\footnotesize
\bibliographystyle{apalike}
\bibliography{references} 

\end{document}